\documentclass[12pt]{article}

\begin{document}

\title{\large {\bf HAWKING RADIATION OF BLACK $p$-BRANES FROM
                        GRAVITATIONAL ANOMALY } }

\author{{\large Zheng Ze Ma}
    \\  \\ {\normalsize \sl Department of Physics, Southeast University,
         Nanjing, 210096, P. R. China }
    \\  {\normalsize \sl z.z.ma@seu.edu.cn } }

\date{}

\maketitle

\vskip 1cm

\begin{minipage}{15cm}

\baselineskip 17 pt

We investigate the Hawking radiation of black $p$-branes of
superstring theories using the method of anomaly cancelation,
specially, we use the method of [S. Iso, H. Umetsu and F. Wilczek,
{\sl Phys. Rev. Lett.} {\bf 96}, 151302 (2006); {\sl Phys. Rev. D}
{\bf 74}, 044017 (2006)]. The metrics of black $p$-branes are
spherically symmetric, but not the Schwarzschild type. In order to
simplify the calculation, we first make a coordinate transformation
to transform the metric to the Schwarzschild type. Then we calculate
its energy-momentum flux from the method of anomaly cancelation of
the above mentioned references. The obtained energy-momentum flux is
equal to a black body radiation, the thermodynamic temperature of
the radiation is equal to its Hawking temperature. And we find that
the results are not changed for the original non-Schwarzschild type
spherically symmetric metric.

\vskip 0.8cm

\noindent {\sl Keywords}: Hawking radiation; anomaly cancelation;
black brane

\vskip 0.8cm

\noindent PACS nos.: 04.50.+h, 04.62.+v, 04.70.Dy, 11.30.-j

\end{minipage}

\newpage

\baselineskip 17 pt

\section{Introduction}

  Hawking radiation is an important physical property of black
holes. Since its original discovery by Hawking,$^{1}$ many people
have tried different methods for the derivation of this
phenomenon.$^{2-6}$ The same result is obtained. Different methods
for the derivation of Hawking radiation show that Hawking radiation
is related with the quantum effect of the gravitational field of a
black hole.

  Recently, a new method for the derivation of Hawking radiation
has been proposed by Robinson and Wilczek {\sl et al}. $^{7,8}$
which is named anomaly cancelation. Robinson and Wilczek {\sl et
al}. $^{7,8}$ find that the effective field theory of quantum fields
near a black hole's horizon is a two-dimensional chiral field theory
due to the fact that a black hole's horizon is a one-way membrane.
Thus there exist gauge and gravitational anomalies for the currents
near a black hole's horizon. However the effective action of quantum
fields near a black hole's horizon is still gauge invariant and
general covariant. Then to combine the regular conditions for the
covariantly anomalous currents on the horizon, gauge and
energy-momentum fluxes with the Hawking temperature $T_{H}$ are
derived outside the horizon. In fact, the germination of this idea
has appeared in a paper of Christensen and Fulling many years
ago,$^{5}$ where the Hawking radiation of a $(1+1)$-dimensional
Schwarzschild black hole was derived from the conformal anomaly
method. The difference lies in that the primal method of Christensen
and Fulling $^{5}$ is only applicable to $(1+1)$-dimensional
spacetime manifold, while the method of Robinson and Wilczek {\sl et
al}. $^{7,8}$ can be applied to higher dimensional spacetime.

  In Refs. 9--11, the method of anomaly cancelation for the
derivation of Hawking radiation has been generalized to
higher-dimensional rotating black holes. Then it has been used to
the derivation of Hawking radiation of various types of black
holes.$^{12-18}$ Some recent developments on this topic have been
carried out in Refs. 19--21. In this paper, we study the Hawking
radiation of general spherically symmetric black $p$-branes of
superstring theories $^{22-26}$ using the method of anomaly
cancelation. The reduced two-dimensional metrics of these black
branes are still spherically symmetric, but not the Schwarzschild
type generally. It is necessary to point out that in the original
work of Robinson and Wilczek {\sl et al},$^{7,8}$ there are some
differences between the method of Ref. 7 and Refs. 8. In Ref. 7, the
Hawking flux of energy-momentum tensor is determined through
canceling the gravitational anomaly in the consistent form at the
horizon. In Refs. 8, the Hawking fluxes are determined by the
conditions that the covariant current and energy-momentum tensor
vanish at the horizon, instead of the consistent current.$^{11}$ For
the Hawking radiation of non-Schwarzschild type spherically
symmetric black holes from the method of anomaly cancelation, some
results have been given by Ref. 17 using the method of Ref. 7.
However, because the method of Refs. 8 is more widely used than the
method of Ref. 7 for the calculation of Hawking fluxes of different
types of black holes, as we can see from Refs. 10--16, etc., it is
necessary for us to give a derivation of the Hawking radiation of
general non-Schwarzschild type spherically symmetric black holes to
use the method of Refs. 8. Although some previous results on this
problem have been given in Ref. 16, in this paper, we will give a
different resolution for this problem.

  In order to derive the Hawking radiation of general non-Schwarzschild
type spherically symmetric black hole metrics from the method of
anomaly cancelation of Refs. 8, we find that it is convenient to
make a coordinate transformation to transform the metric to the form
of the Schwarzschild type. This paper is organized as follows. In
Sec. 2, we briefly review the metrics of black $p$-branes. In Sec.
3, we study the effective action of quantum fields near the horizon
of a black brane. We obtain that it can be described by a
two-dimensional field theory in a curved background. In Sec. 4, we
make a coordinate transformation for the obtained non-Schwarzschild
type two-dimensional spherically symmetric metric to the form of the
Schwarzschild type. In Sec. 5, we calculate the energy-momentum flux
related with the gravitational anomaly for the transformed
two-dimensional metric. We obtain that the energy-momentum flux of a
black $p$-brane obtained from the method of anomaly cancelation
matches its Hawking radiation. Sec. 6 is the conclusion. In Appendix
A, we derive the conditions for the Schwarzschild type coordinate
transformation of Sec. 4.

\section{The Metrics of Black $p$-Branes}

  In this section, we review the black $p$-brane solutions that
appear in superstring theories.$^{22-26}$ We consider the following
$D$-dimensional action
$$
  S=\frac{1}{16\pi}\int d^{D}x\sqrt{-g}\left[ R-\frac{1}{2}
    (\partial\phi)^{2}-\frac{1}{2(d+1)!}e^{-\alpha\phi}
    F^{2}_{d+1}\right],
  \eqno{(1)}  $$
where $\phi$ is a dilaton, $F_{d+1}$ is the field strength for an
antisymmetric tensor $A_{d}$ of rank $d$, i.e.,
$$
  F_{d+1}=dA_{d} ~.
  \eqno{(2)}  $$
This action is a composition part of the low energy effective
actions of type IIA and type IIB superstring theories. It is also a
composition part of the action of eleven-dimensional supergravity,
which is considered as the low energy effective action of M-theory.

  The field equations of the action (1) have the following black
$(d-1)$-brane solution:
\setcounter{equation}{2}
\begin{eqnarray}
  ds^{2} & = & -\left[1-\left(\frac{r_{+}}{r}\right)^{\tilde{d}}
       \right]\left[1-\left(\frac{r_{-}}{r}\right)^{\tilde{d}}
       \right]^{\frac{4\tilde{d}}{\delta(d+\tilde{d})}-1}dt^{2}
       \nonumber    \\
       & + & \left[1-\left(\frac{r_{+}}{r}\right)^{\tilde{d}}
       \right]^{-1}\left[1-\left(\frac{r_{-}}{r}\right)^{\tilde{d}}
       \right]^{\frac{2\alpha^{2}}{\delta\tilde{d}}-1}dr^{2}
       \nonumber    \\
       & + & r^{2}\left[1-\left(\frac{r_{-}}{r}\right)^{\tilde{d}}
       \right]^{\frac{2\alpha^{2}}{\delta\tilde{d}}}d\Omega^{2}
       _{\tilde{d}+1}  \nonumber      \\
       & + & \left[1-\left(\frac{r_{-}}{r}\right)^{\tilde{d}}
       \right]^{\frac{4\tilde{d}}{\delta(d+\tilde{d})}}\delta_{ij}
       dx^{i}dx^{j} ~,
\end{eqnarray}
$$
  A_{01\dots d-1}=\sqrt{\frac{4}{\delta}}\left(\frac{r_{+}r_{-}}
         {r^{2}}\right)^{\frac{\tilde{d}}{2}},
  \eqno{(4)}  $$
$$
  e^{-2\phi}=\left[1-\left(\frac{r_{-}}{r}\right)^{\tilde{d}}
        \right]^{-\frac{4\alpha}{\delta}},
  \eqno{(5)}  $$
where $i,j=1,2,\dots,d-1$, $d\Omega_{n}^{2}$ is the metric of a unit
$n$-sphere. In Eqs. (3)--(5), we have defined
$$
  \tilde{d}=D-d-2 ~,
  \eqno{(6)}  $$
$$
  \alpha^{2}=\delta-\frac{2d(D-d-2)}{D-2} ~.
  \eqno{(7)}  $$
The metric (3) has an event horizon at $r=r_{+}$, and an inner
horizon at $r=r_{-}$, in condition that $r_{+}>r_{-}$. The case of
$r_{+}=r_{-}$ corresponds to the extremal BPS state. In the
following of this paper, we study the non-extremal case of this
metric. We also use $r_{H}$ to represent the radius of the event
horizon in the following.

  The Ramond-Ramond field strength for this solution is given by
$$
  e^{-\alpha\phi}\ast F_{d+1}=\sqrt{\frac{4}{\delta}}\tilde{d}
       (r_{+}r_{-})^{\frac{\tilde{d}}{2}}\epsilon_{\tilde{d}+1} ~,
  \eqno{(8)}  $$
where $\epsilon_{\tilde{d}+1}$ is the volume form on a unit
($\tilde{d}+1$)-sphere. The Ramond-Ramond charge with respect to the
rank $(d+1)$ field strength that the black $(d-1)$-brane carries is
\setcounter{equation}{8}
\begin{eqnarray}
  Q & = & \frac{1}{16\pi}\int\limits_{S^{\tilde{d}+1}}
          e^{-\alpha\phi}\ast F_{d+1}   \nonumber   \\
    & = & \frac{1}{16\pi}\Omega_{\tilde{d}+1}
          \sqrt{\frac{4}{\delta}}\tilde{d}
          (r_{+}r_{-})^{\frac{\tilde{d}}{2}} ~,
\end{eqnarray}
where $\Omega_{n}$ is the volume of a unit $n$-sphere.

  The metrics of general spherically symmetric black branes including
the metric (3) can be written in the form
$$
  ds^{2}=-A(r)dt^{2}+B(r)dr^{2}+r^{2}C(r)d\Omega^{2}_{\tilde{d}+1}
         +D(r)\delta_{ij}dx^{i}dx^{j} ~.
  \eqno{(10)}  $$
Their horizons are determined by $1/B(r)\vert_{r=r_{H}}=0$. The
Hawking temperature for the metric (10) can be obtained from the
Euclidean approach. It is given by
\setcounter{equation}{10}
\begin{eqnarray}
  T_{H} & = & \frac{1}{2\pi}\frac{\partial_{r}\sqrt{A(r)}}
          {\sqrt{B(r)}}\Bigg\vert_{r=r_{H}}  \nonumber    \\
    & = & \frac{1}{4\pi}\frac{A^{\prime}(r)}{\sqrt{A(r)B(r)}}
          \Bigg\vert_{r=r_{H}}~.
\end{eqnarray}
For a general spherically symmetric black hole or black brane
including the metric (3), its horizon is coincident with the surface
of infinite red-shift, thus, $A(r)$ and $B(r)$ can be decomposed
into
$$
  A(r)=a(r)b(r) ~, ~~~~~~~~~
  B(r)=\frac{c(r)}{a(r)} ~,
  \eqno{(12)}  $$
where
$$
  a(r_{H})=0
  \eqno{(13)}  $$
determines the radius of the horizon, as well as the radius of the
surface of infinite red-shift; but $b(r_{H})\neq 0$, $c(r_{H})\neq
0$. Here, we have $A(r)B(r)=b(r)c(r)\neq 1$ generally.

  To substitute Eqs. (12) in Eq. (11), and to consider Eq. (13), the
term that comes from the derivative of $b(r)$ vanishes. This yields
$$
  T_{H}=\frac{1}{4\pi}\left(a^{\prime}(r)\sqrt{\frac{b(r)}{c(r)}} ~
        \right)\Bigg\vert_{r=r_{H}} ~.
  \eqno{(14)}  $$
For the metric (3), we have
\setcounter{equation}{14}
\begin{eqnarray}
  a(r) & = & 1-\left(\frac{r_{+}}{r}\right)^{\tilde{d}} ~,
        \nonumber    \\
  b(r) & = & \left[1-\left(\frac{r_{-}}{r}\right)^{\tilde{d}}
       \right]^{\frac{4\tilde{d}}{\delta(d+\tilde{d})}-1},
        \nonumber    \\
  c(r) & = & \left[1-\left(\frac{r_{-}}{r}\right)^{\tilde{d}}
       \right]^{\frac{2\alpha^{2}}{\delta\tilde{d}}-1}.
\end{eqnarray}
To substitute Eqs. (15) in Eq. (14), or from Eq. (11), we obtain,
for the metric (3),
$$
  T_{H}=\frac{\tilde{d}}{4\pi r_{+}}\left[1-\left(
          \frac{r_{-}}{r_{+}}\right)^{\tilde{d}}\right]
          ^{\frac{2\tilde{d}}{\delta(d+\tilde{d})}-
            \frac{\alpha^{2}}{\delta\tilde{d}}}.
  \eqno{(16)}  $$
To make use of Eqs. (6) and (7), we obtain, at last,
$$
  T_{H}=\frac{\tilde{d}}{4\pi r_{+}}\left[1-\left(
          \frac{r_{-}}{r_{+}}\right)^{\tilde{d}}\right]
          ^{\frac{2}{\delta}-\frac{1}{\tilde{d}}}.
  \eqno{(17)}  $$
The chemical potential with respect to the Ramond-Ramond charge is
the value of the gauge potential at the horizon, i.e.,
$$
  \mu=A_{01\dots d-1}\vert_{r=r_{H}}=\sqrt{\frac{4}{\delta}}
      \left(\frac{r_{-}}{r_{+}}\right)^{\frac{\tilde{d}}{2}}.
  \eqno{(18)}  $$
The first law of thermodynamics is satisfied by the black
$(d-1)$-branes.

\section{Quantum Fields Near The Horizon}

  Robinson and Wilczek {\sl et al}. have shown that quantum fields
near a black hole horizon can be described by a two-dimensional
field theory in a curved background of the Schwarzschild
type.$^{7,8}$ This is the primal recipe for the derivation of
Hawking radiation from the method of anomaly cancelation. Thus, in
order to study the Hawking radiation of black $p$-branes from the
method of anomaly cancelation, we first need to study the effective
field theory of quantum fields near a black brane horizon.

  We consider a free scalar field $\varphi$ in the background of
the metric (10). To be simple, we suppose that the scalar field is
zero-mass, and we suppose that it does not couple with the
Ramond-Ramond gauge field. Then, the action of this system is given
by \setcounter{equation}{18}
\begin{eqnarray}
  S_{{\footnotesize \mbox{free}}}(\varphi)
      & = & \int d^{D}x\sqrt{-g} ~ g^{\mu\nu}
      \partial_{\mu}\varphi\partial_{\nu}\varphi
      \nonumber   \\
      & = & \int d^{D}x\sqrt{-g}\left[\partial_{\mu}(\varphi
        \partial^{\mu}\varphi)-\varphi\partial_{\mu}
        \partial^{\mu}\varphi\right] ~,
\end{eqnarray}
where $g$ is the determinant of the metric (10), and thus
$$
  \sqrt{-g}=r^{\tilde{d}+1}\sqrt{A(r)B(r)C^{\tilde{d}+1}(r)
            D^{d-1}(r)} ~.
  \eqno{(20)}  $$
In considering that $\sqrt{-g}$ only depends on the coordinate $r$,
while does not depend on the other coordinates, to omit a surface
term in the action, we obtain
$$
  S_{{\footnotesize \mbox{free}}}(\varphi)=
     -\int d^{D}x\sqrt{-g}\varphi(\partial_{\mu}
      \partial^{\mu}+\frac{1}{\sqrt{-g}}
      \partial_{r}\sqrt{-g}\partial^{r})\varphi ~.
  \eqno{(21)}  $$
To take the near horizon limit $r\rightarrow r_{H}$, the second term
of Eq. (21) can be omitted compared with the first term. This fact
can be made clear through transforming to the radial ``tortoise"
coordinate as that of Refs. 7 and 8. On the other hand, in the near
horizon limit, $\sqrt{-g}$ tends to a constant, thus, it can be
moved outside the integral. Therefore, near the black brane horizon,
the action is dominated by
$$
  S_{{\footnotesize \mbox{free}}}(\varphi)=
        -\sqrt{-g}\vert_{r_{H}}\int dtdr d\Omega_{\tilde{d}+1}
        d^{d-1}x ~ \varphi
       (\partial_{t}\partial^{t}+\partial_{r}\partial^{r}+
        \partial_{[D-2]}\partial^{[D-2]})\varphi ~,
  \eqno{(22)}  $$
where $d\Omega_{\tilde{d}+1}$ is the volume element of the
$(\tilde{d}+1)$-dimensional unit sphere, $d^{d-1}x$ is the volume
element of the $(d-1)$-dimensional transverse space. Here, we use
$\partial_{[D-2]}\partial^{[D-2]}$ to represent the full Laplacian
on the $(D-2)$-dimensional space whose explicit form is given by
$$
  \partial_{[D-2]}\partial^{[D-2]}=\frac{1}{r^{2}C(r)}
        \nabla^{2}_{\Omega}+\frac{1}{D(r)}\nabla^{2}_{X} ~,
  \eqno{(23)}  $$
where $\nabla^{2}_{\Omega}$ is the Laplacian on the
$(\tilde{d}+1)$-dimensional unit sphere, $\nabla^{2}_{X}$ is the
Laplacian on the $(d-1)$-dimensional transverse space.

  We can expand $\varphi(x)$ in terms of the normalized
eigenfunctions of the operator $\partial_{[D-2]}\partial^{[D-2]}$.
Therefore we have
$$
  \varphi(x)=\sum_{l_{1},\dots,l_{\tilde{d}+1},k_{1},\dots,k_{d-1}}
     \varphi_{l_{1}\dots l_{\tilde{d}+1}k_{1}\dots k_{d-1}}(r,t)
     Y_{l_{1}\dots l_{\tilde{d}+1}}(\theta_{1},\dots,
     \theta_{\tilde{d}+1})X(k_{1},\dots,k_{d-1}) ~.
  \eqno{(24)}  $$
In Eq. (24), $Y_{l_{1}\dots l_{\tilde{d}+1}}(\theta_{1},\dots,
\theta_{\tilde{d}+1})$ are the normalized spherical harmonics on a
$(\tilde{d}+1)$-dimensional unit sphere with the azimuthal angles
$(\theta_{1},\dots, \theta_{\tilde{d}+1})$, and
$$
  X(k_{1},\dots,k_{d-1})=e^{i(k_{1}x^{1}+
          \cdots+k_{d-1}x^{d-1})} ~,
  \eqno{(25)}  $$
they satisfy
\setcounter{equation}{25}
\begin{eqnarray}
  & ~ &     \!\!\!\!\!\!\!\!\!\!\!\!\!\!\!\!
   \int d\Omega_{\tilde{d}+1}d^{d-1}x
        Y_{l_{1}\dots l_{\tilde{d}+1}}(\theta_{1},\dots,
       \theta_{\tilde{d}+1})X(k_{1},\dots,k_{d-1})\times
       \nonumber   \\
  & ~ &
       Y_{m_{1}\dots m_{\tilde{d}+1}}(\theta_{1},\dots,
       \theta_{\tilde{d}+1})X^{\ast}(p_{1},\dots,p_{d-1})
       =\delta_{l_{1}m_{1}}\cdots \delta_{l_{\tilde{d}+1}
       m_{\tilde{d}+1}}\delta_{k_{1}p_{1}}\cdots
       \delta_{k_{d-1}p_{d-1}} ~,
\end{eqnarray}
where the integration of the coordinates $x^{i}$ takes a unit volume
on the transverse space of the brane.

  In Eq. (24),
$\varphi_{l_{1}\dots l_{\tilde{d}+1}k_{1}\dots k_{d-1}}(r,t)$ and
$Y_{l_{1}\dots l_{\tilde{d}+1}}(\theta_{1},\dots,
\theta_{\tilde{d}+1})$ are real functions. However, because
$X(k_{1},\dots,k_{d-1})$ given by Eq. (25) are complex functions,
$\varphi(x)$ given by Eq. (24) has been complexified. This means
that in Eq. (22) inside the integral, the first $\varphi$ should be
replaced by $\varphi^{\ast}$. Thus, to substitute Eq. (24) in Eq.
(22), the near horizon effective action of the free scalar field is
obtained as
$$
  S_{{\footnotesize \mbox{free}}}(\varphi)=
       -\sqrt{-g}\vert_{r_{H}}\!\!\sum_{l_{1},\dots,
       l_{\tilde{d}+1},k_{1},\dots,k_{d-1}}\!\!\int dtdr
       \varphi_{l_{1}\dots
     l_{\tilde{d}+1}k_{1}\dots k_{d-1}}(r,t)
    (\partial_{t}\partial^{t}+\partial_{r}\partial^{r})
     \varphi_{l_{1}\dots l_{\tilde{d}+1}k_{1}\dots k_{d-1}}(r,t) ~.
  \eqno{(27)}  $$
From Eq. (27), we can see that, near the black brane horizon, the
effective field theory of the free scalar field is a two-dimensional
field theory in a curved background with the metric
$$
  ds^{2}=-A(r)dt^{2}+B(r)dr^{2} ~.
  \eqno{(28)}  $$
The metric (28) is spherically symmetric, but not the Schwarzschild
type generally, because $A(r)B(r)\neq 1$ generally. We can expect
that for the other fields near the black brane horizon, their
effective field theories are also two-dimensional field theories in
a curved background of the metric (28). The explicit form of the
reduced two-dimensional metric of the metric (3) is given by
\setcounter{equation}{28}
\begin{eqnarray}
  ds^{2} = & - & \left[1-\left(\frac{r_{+}}{r}\right)^{\tilde{d}}
       \right]\left[1-\left(\frac{r_{-}}{r}\right)^{\tilde{d}}
       \right]^{\frac{4\tilde{d}}{\delta(d+\tilde{d})}-1}dt^{2}
       \nonumber    \\
       & + & \left[1-\left(\frac{r_{+}}{r}\right)^{\tilde{d}}
       \right]^{-1}\left[1-\left(\frac{r_{-}}{r}\right)^{\tilde{d}}
       \right]^{\frac{2\alpha^{2}}{\delta\tilde{d}}-1}dr^{2} ~.
\end{eqnarray}

  If the scalar field carries the Ramond-Ramond charge, then it will
couple with the Ramond-Ramond gauge field of the black brane.
However, because the explicit forms of the covariant couplings of
quantum fields with the Ramond-Ramond gauge fields are not clear, we
do not consider such a coupling here in this paper; therefore, we
will not study the Ramond-Ramond charge flux related with the gauge
anomaly in the following.

\section{Schwarzschild Type Coordinate Transformation}

  To use the metric (28) directly to calculate the Hawking flux
using the method of Refs. 8 is not convenient for the case
$A(r)B(r)\neq 1$. In order to simplify the calculation, we can make
a coordinate transformation to transform the metric (28) to the
Schwarzschild type first. They relate with the coordinates $(t,r)$
through the transformation
$$
  t=t(\tau,r^{\ast}) ~, ~~~~~~~~~~
  r=r(\tau,r^{\ast}) ~.
  \eqno{(30)}  $$
We demand that under this coordinate transformation, the metric (28)
transforms to the form
$$
  ds^{2}=-\sqrt{\frac{A(r^{\ast})}{B(r^{\ast})}}d\tau^{2}+
         \sqrt{\frac{B(r^{\ast})}{A(r^{\ast})}}dr^{\ast 2} ~,
  \eqno{(31)}  $$
which is now the Schwarzschild type. Here, we mean that the
expression of the function $A(r^{\ast})$ is the same as the
expression of the function $A(r)$, the expression of the function
$B(r^{\ast})$ is the same as the expression of the function $B(r)$.
The partial differential equations that the transformation (30)
should satisfy are derived in Appendix A. Here, we only point out
that such a kind of coordinate transformation is existing, it is
also not unique; however, we need not to obtain its explicit form.

  We can write Eq. (31) in the following form:
$$
  ds^{2}=-F(r^{\ast})d\tau^{2}+\frac{1}{F(r^{\ast})}dr^{\ast 2} ~,
  \eqno{(32)}  $$
where
$$
  F(r^{\ast})=\sqrt{\frac{A(r^{\ast})}{B(r^{\ast})}} ~.
  \eqno{(33)}  $$
As postulated above, the expression of the function $A(r^{\ast})$ is
the same as the expression of the function $A(r)$, the expression of
the function $B(r^{\ast})$ is the same as the expression of the
function $B(r)$. Thus, according to Eqs. (12), the functions
$A(r^{\ast})$ and $B(r^{\ast})$ can be decomposed into the form
$$
  A(r^{\ast})=a(r^{\ast})b(r^{\ast}) ~, ~~~~~~~~~
  B(r^{\ast})=\frac{c(r^{\ast})}{a(r^{\ast})} ~,
  \eqno{(34)}  $$
where the expressions of the functions $a(r^{\ast})$, $b(r^{\ast})$,
and $c(r^{\ast})$ are the same as the expressions of the functions
$a(r)$, $b(r)$, and $c(r)$ respectively. To substitute Eqs. (34) in
Eq. (33), we can write
$$
  F(r^{\ast})=a(r^{\ast})\sqrt{\frac{b(r^{\ast})}{c(r^{\ast})}} ~.
  \eqno{(35)}  $$
Then the condition
$$
  a(r^{\ast}_{H})=0
  \eqno{(36)}  $$
determines the radius of the horizon of the metric (31) or (32), but
$b(r^{\ast}_{H})\neq 0$, $c(r^{\ast}_{H})\neq 0$. However, because
the expression of the function $a(r^{\ast})$ is the same as the
expression of the function $a(r)$, from Eqs. (36) and (13), we have
$$
  r^{\ast}_{H}=r_{H} ~,
  \eqno{(37)}  $$
which means that the location of the horizon of the metric (31) or
(32) is the same as the location of the horizon of the metric (28),
i.e., the location of the horizon is not changed after the
Schwarzschild type coordinate transformation (30).

  For the Schwarzschild type metric (32), its Hawking
temperature is given by
$$
  T^{\ast}_{H}=\frac{1}{4\pi}F^{\prime}(r^{\ast})
               \vert_{r^{\ast}=r^{\ast}_{H}} ~.
  \eqno{(38)}  $$
To substitute Eq. (35) in Eq. (38), and to consider that on the
horizon $a(r^{\ast}_{H})=0$, we obtain
$$
  T^{\ast}_{H}=\frac{1}{4\pi}\left(a^{\prime}(r^{\ast})
        \sqrt{\frac{b(r^{\ast})}{c(r^{\ast})}} ~
        \right)\Bigg\vert_{r^{\ast}=r^{\ast}_{H}} ~.
  \eqno{(39)}  $$
As mentioned above, the expressions of the functions $a(r^{\ast})$,
$b(r^{\ast})$, and $c(r^{\ast})$ are the same as the expressions of
the functions $a(r)$, $b(r)$, and $c(r)$ respectively, and because
of Eq. (37), to compare Eq. (39) with Eq. (14), we have
$$
  T_{H}=T^{\ast}_{H} ~,
  \eqno{(40)}  $$
which means that the Hawking temperature of the metric (32) is the
same as the Hawking temperature of the metric (28). The Hawking
temperature of the metric (28) is not changed after the coordinate
transformation (30). For the two-dimensional metric of Eq. (29),
when transformed to the form of Eq. (32), its explicit form is given
by
\setcounter{equation}{40}
\begin{eqnarray}
  ds^{2} & = & -\left[1-\left(\frac{r_{+}}{r^{\ast}}
       \right)^{\tilde{d}}\right]
       \left[1-\left(\frac{r_{-}}{r^{\ast}}\right)^{\tilde{d}}
       \right]^{\frac{2\tilde{d}}{\delta(d+\tilde{d})}-
       \frac{\alpha^{2}}{\delta\tilde{d}}}d\tau^{2}
       \nonumber    \\
       & + & \left[1-\left(\frac{r_{+}}{r^{\ast}}
       \right)^{\tilde{d}}\right]^{-1}
       \left[1-\left(\frac{r_{-}}{r^{\ast}}\right)^{\tilde{d}}
       \right]^{\frac{\alpha^{2}}{\delta\tilde{d}}-
       \frac{2\tilde{d}}{\delta(d+\tilde{d})}}dr^{\ast 2} ~.
\end{eqnarray}
Here, the radius of the event horizon is $r^{\ast}_{H}=r_{+}$. It is
not changed under the coordinate transformation (30). The radius of
the inner horizon is $r_{-}$. It is not changed under the coordinate
transformation (30) either. These facts can be seen from the metric
(41). For the metric (41), its Hawking temperature can be obtained
from Eq. (38) or (39) which is
$$
  T^{\ast}_{H}=\frac{\tilde{d}}{4\pi r_{+}}\left[1-\left(
          \frac{r_{-}}{r_{+}}\right)^{\tilde{d}}\right]
          ^{\frac{2}{\delta}-\frac{1}{\tilde{d}}}.
  \eqno{(42)}  $$
It is just equal to the Hawking temperature of the metric (29) given
by Eq. (17).

  From the above analysis, we can see that, to perform a
coordinate transformation (30), the metric (28) can be transformed
to the form of Eq. (31) or (32), which is the Schwarzschild type.
For a black hole or a black brane, because it is a thermal
equilibrium system, its Hawking radiation only relies on its Hawking
temperature and horizon's location in fact. However, for the metric
(31) or (32), we have seen that its Hawking temperature and
horizon's location are just the same as the Hawking temperature and
horizon's location of the metric (28). This means that we can use
the metric (31) or (32) equivalently to calculate the Hawking flux
for the metric (28) using the method of anomaly cancelation.

\section{Gravitational Anomaly And Energy-momentum Flux}

  In this section, we calculate the energy-momentum flux for the
metric (32) using the method of anomaly cancelation. We will turn
back to the metric (28) equivalently at last. Following Refs. 8, to
consider the area outside the horizon, we divide it into two parts:
$[r^{\ast}_{H},r^{\ast}_{H}+\epsilon]$ and
$[r^{\ast}_{H}+\epsilon,\infty]$.
$[r^{\ast}_{H},r^{\ast}_{H}+\epsilon]$ is the near horizon part,
where the physics has certain exotic properties.
$[r^{\ast}_{H}+\epsilon,\infty]$ is the part departed from the
horizon, where the physics has the usual properties. The parameter
$\epsilon$ can be taken arbitrarily small, thus, for the observably
physical results, we can take them in the region
$[r^{\ast}_{H}+\epsilon,\infty]$ always.

  In the near horizon region $[r^{\ast}_{H},r^{\ast}_{H}+\epsilon]$,
to consider that a black hole's horizon is a one-way membrane, for
the $(1+1)$-dimensional field theory, the ingoing (left moving)
modes will tend to the center singularity, hence they will not
affect the physics of the region
$[r^{\ast}_{H},r^{\ast}_{H}+\epsilon]$. That is to say in the region
$[r^{\ast}_{H},r^{\ast}_{H}+\epsilon]$, only the outgoing (right
moving) modes are responsible for the observable physics. This makes
the practical field theory be a two-dimensional chiral field theory
in the near horizon region. Thus, in the region
$[r^{\ast}_{H},r^{\ast}_{H}+\epsilon]$, there exists the
gravitational anomaly for the energy-momentum current. In the region
$[r^{\ast}_{H}+\epsilon,\infty]$, the ingoing and outgoing modes are
both existing, the field theory is a normal one while not chiral,
there is no gravitational anomaly for the energy-momentum current.

  For a black hole or a black brane, it is a thermodynamical
equilibrium system, all currents in the spacetime outside the
horizon are static. The energy-momentum tensor outside the horizon
can be decomposed into the form
$$
  T_{\nu}^{\mu}(r^{\ast})=T^{\mu}_{\nu(H)}(r^{\ast})H(r^{\ast})+
          T^{\mu}_{\nu(o)}(r^{\ast})\Theta_{+}(r^{\ast}) ~,
  \eqno{(43)}  $$
where $\Theta_{+}(r^{\ast})=\Theta(r^{\ast}-r_{+}-\epsilon)$ (here
we use $r_{+}$ to represent the radius of the event horizon), and
$H(r^{\ast})=1-\Theta_{+}(r^{\ast})$. Thus,
$T^{\mu}_{\nu(H)}(r^{\ast})$ is the energy-momentum tensor in the
region $[r^{\ast}_{H},r^{\ast}_{H}+\epsilon]$,
$T^{\mu}_{\nu(o)}(r^{\ast})$ is the energy-momentum tensor in the
region $[r^{\ast}_{H}+\epsilon,\infty]$.

  In a two-dimensional spacetime, $T_{t}^{r}(r^{\ast})$ is just the
energy-momentum flux in the radial direction. For
$T_{\nu(o)}^{\mu}(r^{\ast})$, as analyzed above, there is no
gravitational anomaly, it satisfies the normal conservation
equation, therefore we have
$$
  \partial_{r^{\ast}}T_{t(o)}^{r}(r^{\ast})=0 ~.
  \eqno{(44)}  $$
But for $T_{\nu(H)}^{\mu}(r^{\ast})$ near the horizon, it has the
gravitational anomaly. It satisfies the anomalous conservation
equation.$^{7,8,27,28}$ This results
$$
  \partial_{r^{\ast}}T_{t(H)}^{r}(r^{\ast})=
     \partial_{r^{\ast}}N^{r}_{t}(r^{\ast}) ~,
  \eqno{(45)}  $$
where the right hand side of Eq. (45) comes from the gravitational
anomaly of the consistent energy-momentum tensor. For the
two-dimensional metric (32), to use the result of Refs. 7 and 8, we
have
$$
  N^{r}_{t}(r^{\ast})=\frac{1}{192\pi}\left[(F^{\prime}
       (r^{\ast}))^{2}+F^{\prime\prime}(r^{\ast})F(r^{\ast})\right].
  \eqno{(46)}  $$
The integration of Eqs. (44) and (45) yields
\setcounter{equation}{46}
\begin{eqnarray}
  & ~ & T^{r}_{t(o)}(r^{\ast})=a^{\ast}_{o} ~,  \nonumber      \\
  & ~ & T^{r}_{t(H)}(r^{\ast})=a^{\ast}_{H}+
        \int_{r^{\ast}_{H}}^{r^{\ast}}dr^{\ast}
        \partial_{r^{\ast}}N_{t}^{r}(r^{\ast}) ~,
\end{eqnarray}
where $a^{\ast}_{o}$, $a^{\ast}_{H}$ are two integration constants.
From Eq. (43), we can see that $a^{\ast}_{o}$ is just the
energy-momentum flux for an observer to measure outside the horizon.

  On the other hand, the anomaly of the energy-momentum tensor is
purely a quantum field effect, it does not affect the general
covariance of the effective action. To perform an infinitesimal
coordinate transformation for the two-dimensional field theory along
the time direction with the parameter $\xi^{t}$, we obtain $^{8}$
\setcounter{equation}{47}
\begin{eqnarray}
  \delta W
  & = & -\int d^{2}x\sqrt{-g} ~ \xi^{t}
      \nabla_{\mu}T_{t}^{\mu}      \nonumber    \\
  & = & -\int d^{2}x ~ \xi^{t}\left[\partial_{r}(N_{t}^{r}H)
        +\left(T^{r}_{t(o)}-T^{r}_{t(H)}
        +N_{t}^{r}\right)\delta(r^{\ast}-r_{+}-\epsilon)\right] ~.
\end{eqnarray}
As pointed out in Refs. 8, the first term in the second line of Eq.
(48) can be canceled by the quantum effect of the ingoing modes near
the horizon. Thus, general covariance of the effective action leads
the vanishing of the coefficient of the $\delta$-function. Then, to
combine Eqs. (47), we obtain
$$
  a^{\ast}_{o}=a^{\ast}_{H}-N_{t}^{r}(r^{\ast}_{H}) ~.
  \eqno{(49)}  $$
Substituting Eq. (35) in Eq. (46), and considering that on the
horizon $a(r^{\ast}_{H})=0$, we obtain
$$
  N^{r}_{t}(r^{\ast}_{H})=\frac{1}{192\pi}
        \left(a^{\prime}(r^{\ast})\sqrt{\frac{b(r^{\ast})}
        {c(r^{\ast})}} ~ \right)^{2}
        \Bigg\vert_{r^{\ast}=r^{\ast}_{H}} ~.
  \eqno{(50)}  $$

  In order to determine the constant $a^{\ast}_{H}$ of Eq. (49), we
need to introduce the covariantly anomalous energy-momentum tensor
$\widetilde{T}_{\mu\nu}$, it satisfies the covariant anomaly
equation$^{28}$
$$
  \nabla^{\mu}\widetilde{T}_{\mu\nu}=\frac{1}{96\pi\sqrt{-g}}
     \epsilon_{\mu\nu}\partial^{\mu}R ~,
  \eqno{(51)}  $$
where $R$ is the Ricci scalar. For the component
$\widetilde{T}^{r}_{t}$ which is necessary for the following
calculation, for the metric (32), to use the result of Refs. 8, we
have
$$
  \widetilde{T}^{r}_{t}(r^{\ast})=T^{r}_{t}(r^{\ast})+
       \frac{1}{192\pi}\left[F(r^{\ast})F^{\prime\prime}(r^{\ast})
       -2(F^{\prime}(r^{\ast}))^{2}\right].
  \eqno{(52)}  $$
As postulated in Refs. 8, $\widetilde{T}^{r}_{t}$ satisfies the
regular boundary condition
$$
  \widetilde{T}^{r}_{t}(r^{\ast}_{H})=0 ~,
  \eqno{(53)}  $$
such a boundary condition makes physical quantities regular on the
future horizon for a free falling observer.$^{8,9}$ The combination
of Eqs. (52), (53), and (43) yields
$$
  T^{r}_{t(H)}(r^{\ast}_{H})=\frac{1}{192\pi}
      \left[2(F^{\prime}(r^{\ast}))^{2}
      -F(r^{\ast})F^{\prime\prime}(r^{\ast})\right]
      \Big\vert_{r^{\ast}=r^{\ast}_{H}} ~.
  \eqno{(54)}  $$
Substituting for $F(r^{\ast})$ from Eq. (35), and considering that
on the horizon $a(r^{\ast}_{H})=0$, we obtain
$$
  T^{r}_{t(H)}(r^{\ast}_{H})=\frac{1}{96\pi}
        \left(a^{\prime}(r^{\ast})\sqrt{\frac{b(r^{\ast})}
        {c(r^{\ast})}} ~ \right)^{2}
        \Bigg\vert_{r^{\ast}=r^{\ast}_{H}} ~.
  \eqno{(55)}  $$
Then, the combination of Eqs. (55) and (47) results
$$
  a^{\ast}_{H}=\frac{1}{96\pi}
        \left(a^{\prime}(r^{\ast})\sqrt{\frac{b(r^{\ast})}
        {c(r^{\ast})}} ~ \right)^{2}
        \Bigg\vert_{r^{\ast}=r^{\ast}_{H}} ~.
  \eqno{(56)}  $$
Substituting Eqs. (56) and (50) in Eq. (49), we obtain
$$
  a^{\ast}_{o}=\frac{1}{192\pi}
        \left(a^{\prime}(r^{\ast})\sqrt{\frac{b(r^{\ast})}
        {c(r^{\ast})}} ~ \right)^{2}
        \Bigg\vert_{r^{\ast}=r^{\ast}_{H}} ~.
  \eqno{(57)}  $$
Finally, comparing Eq. (57) with Eq. (39), we can write
$$
  a^{\ast}_{o}=\frac{\pi}{12}T^{\ast 2}_{H}  ~,
  \eqno{(58)}  $$
where $T^{\ast}_{H}$ is just the Hawking temperature of the metric
(32). The explicit expression of $T^{\ast}_{H}$ of the metric (41)
is given by Eq. (42). As mentioned above, $a^{\ast}_{o}$ is just the
energy-momentum flux for an observer to measure outside the horizon
of the metric (32). From the above derivation, we can see that the
anomalous current $a^{\ast}_{H}$ near the horizon, i.e., the
outgoing chiral current, has contribution to the current
$a^{\ast}_{o}$ departed from the horizon which is a normal one, or,
we can say that the existence of the outgoing chiral current makes
the radiation current be a normal one and cancels its anomaly.

  To consider a two-dimensional black body radiation with
temperature $T$, the distribution of a zero-mass bose field is given
by
$$
  N(\omega)=\frac{1}{e^{\omega/T}-1} ~.
  \eqno{(59)}  $$
Here, we need not to consider the distribution of a fermion field in
order to avoid the superradiance problem, because there is no
superradiance for a non-rotating black hole or black brane. The
energy-momentum flux of this two-dimensional black body radiation is
obtained as
$$
  F_{E}=\frac{1}{2\pi}\int_{0}^{\infty}\omega N(\omega)d\omega
        =\frac{\pi}{12}T^{2} ~.
  \eqno{(60)}  $$
To compare Eq. (58) with Eq. (60), we can see that Eq. (58) is just
the energy-momentum flux of a two-dimensional black body radiation
with the temperature $T^{\ast}_{H}$. Therefore, for the transformed
two-dimensional Schwarzschild type metric (32), it has a black body
radiation with the thermal temperature given by its Hawking
temperature.

  For a black hole or a black brane, to be a thermal equilibrium
system, its Hawking radiation should only be determined by its
Hawking temperature and horizon's location. For the metric (28), its
Hawking temperature and horizon's location are just the same as the
Hawking temperature and horizon's location of the metric (32), its
Hawking temperature and horizon's location are not changed under the
coordinate transformation (30), thus, we can deduce that, for the
metric (28), it has a Hawking radiation as same as the Hawking
radiation of the metric (32). This means that for the original
non-Schwarzschild type spherically symmetric metric (28), we have,
outside its horizon, a radial energy-momentum flux given by
$$
  a_{o}=\frac{\pi}{12}T^{2}_{H}  ~,
  \eqno{(61)}  $$
where $T_{H}$ is just its Hawking temperature. Hence, we have
derived that there is a Hawking radiation for the two-dimensional
non-Schwarzschild type spherically symmetric metric (28) outside its
horizon from the method of anomaly cancelation of Refs. 8, and the
temperature of the thermal radiation derived from the method of
anomaly cancelation is in accordance with its Hawking temperature
derived from the black brane thermodynamics.

  There is a problem needing to be deliberated further. From Eqs.
(47), $a^{\ast}_{o}=T^{r}_{t(o)}(r^{\ast})$, at the same time,
$T^{r}_{t(o)}(r^{\ast})$ is a component of the second-order tensor
$T^{\mu}_{\nu(o)}(r^{\ast})$. For the energy-momentum flux
$T^{r}_{t(o)}(r)$ of the original metric (28), which is a component
of the second-order tensor $T^{\mu}_{\nu(o)}(r)$ of the original
metric (28), we need to obtain it from the coordinate transformation
(30) in principle, and it will not remain to be a constant of
$a^{\ast}_{o}$ generally because of the nonlinearity of the
coordinate transformation. Thus it seems that we can not obtain the
correct result for the Hawking flux of the original metric (28).
However, such a problem is only superficial, it comes from the
coordinate transformation in fact. This is because, first, for the
metrics (28) and (32), or (29) and (41), they are all asymptotic
flat. Therefore, at spacelike infinity, the coordinate
transformation (30) and its inverse transformation will tend to an
identical transformation necessarily. This means that for the
original metric (28), at infinity, we have
$T^{r}_{t(o)}(\infty)=a^{\ast}_{o}$. Next, for the tensor
$T^{\mu}_{\nu(o)}(r)$ of the original metric (28), there is no
gravitational anomaly, like that of Eq. (44), it satisfies the
normal conservation equation. This results
$$
  \partial_{r}T_{t(o)}^{r}(r)=0 ~,
  \eqno{(62)}  $$
whose solution is given by
$$
  T_{t(o)}^{r}(r)=a_{o} ~,
  \eqno{(63)}  $$
where $a_{o}$ is a constant. But we have known from the above
analysis that, at infinity, $T^{r}_{t(o)}(\infty)=a^{\ast}_{o}$.
Then, from Eq. (63), we have
$$
  T_{t(o)}^{r}(r)=a^{\ast}_{o}
  \eqno{(64)}  $$
for an arbitrary $r$. Because $a^{\ast}_{o}$ has been obtained in
Eq. (58), we obtain, for the original metric (28), its
energy-momentum flux is given by
$$
  a_{o}=\frac{\pi}{12}T^{\ast 2}_{H} ~.
  \eqno{(65)}  $$
In considering Eq. (40), we can write
$$
  a_{o}=\frac{\pi}{12}T^{2}_{H} ~,
  \eqno{(66)}  $$
where $T_{H}$ is the Hawking temperature of the original metric
(28). This means that the energy-momentum flux of the original
two-dimensional metric (28) is a constant, and, it is equal to a
two-dimensional black body radiation, the temperature of the
radiation is just its Hawking temperature obtained from the black
brane thermodynamics. For the original higher-dimensional metric (3)
or (10), from the mode decomposition of the field in terms of Eq.
(24), it is not difficult to see that the distribution of the
spectrum will not change. Therefore, for the black brane metric (3),
we can conclude that it has a Hawking radiation with the temperature
$T_{H}$ from the method of anomaly cancelation.

  However, if we do not perform the coordinate transformation (30)
and use the original metric (28) directly to calculate the
energy-momentum flux from the method of anomaly cancelation, then,
for $\widetilde{T}^{r}_{t}(r)$, which is a component of the
covariantly anomalous energy-momentum tensor
$\widetilde{T}_{\nu}^{\mu}(r)$, it seems that there will be some
additional factors of the form $\sqrt{-g_{2}}=\sqrt{A(r)B(r)}$, and
they can not be canceled due to the metric is not the Schwarzschild
type. Therefore, we constructed a coordinate transformation (30) to
transform the metric to the Schwarzschild type first, and then, we
use the method of anomaly cancelation of Refs. 8 to calculate its
Hawking flux.

\section{Conclusion}

  In this paper, we studied the Hawking radiation of black $p$-branes
of superstring theories $^{22-26}$ from the method of anomaly
cancelation of Refs. 8. The metrics of these black $p$-branes are
spherically symmetric, but not the Schwarzschild type. The
calculation of this paper are carried out with respect to the
general non-Schwarzschild type spherically symmetric black hole
metrics.

  Some previous works have been taken for the Hawking fluxes of
non-Schwarzschild type spherically symmetric black hole metrics
using the method of anomaly cancelation.$^{16-18}$ In Ref. 18,
Hawking radiation of the $D1$-$D5$ brane black holes of superstring
theories have been calculated using the method of anomaly
cancelation. The black hole metrics studied in Ref. 18 are
non-Schwarzschild type; however, they belong to certain special kind
of the metrics of general non-Schwarzschild type spherically
symmetric black holes.

  In Ref. 17, Hawking radiation are also calculated with respect to
the general non-Schwarzschild type spherically symmetric black hole
metrics. However, the method adopted in this paper is different from
that of Ref. 17. In Ref. 17, Hawking radiation are calculated using
the method of Ref. 7, i.e., the Hawking flux of energy-momentum
tensor is determined through canceling the gravitational anomaly in
the consistent form at the horizon. In this paper, we used the
method of Refs. 8, i.e., the Hawking fluxes are determined by the
conditions that the covariant current and energy-momentum tensor
vanish at the horizon, instead of the consistent current. Because
the method of Refs. 8 is more general than the method of Ref. 7 for
the calculation of Hawking fluxes of different types of black holes,
as we can see from Refs. 10--16, etc., it is necessary for us to
study the Hawking radiation of general non-Schwarzschild type
spherically symmetric black hole metrics to use the method of Refs.
8.

  On the other hand, we can see that in Ref. 16, Hawking radiation
of general non-Schwarzschild type spherically symmetric black hole
metrics from the method of Refs. 8 has been analyzed to use the
metric (28) directly, while not through transforming it to the
Schwarzschild type. However, many formulas have become complicated
because of this as we can see from Ref. 16. In order to give a
clearer derivation for this problem, we find that it is convenient
to make a coordinate transformation to transform the
non-Schwarzschild type spherically symmetric black hole metric to
the Schwarzschild type first, and then to calculate the
energy-momentum flux for the transformed metric from the method of
anomaly cancelation. The obtained energy-momentum flux is equal to a
black body radiation, the thermodynamic temperature of the radiation
is equal to the Hawking temperature. And we find that the results
are not changed for the original non-Schwarzschild type spherically
symmetric metric. Then, from the mode decomposition of the scalar
field in terms of Eq. (24), we can deduce that the distribution of
the spectrum is not changed for the whole higher-dimensional object.
Thus, we have derived the Hawking radiation of black $p$-branes from
the method of anomaly cancelation of Refs. 8. The result of this
paper is held for a general higher-dimensional non-Schwarzschild
type spherically symmetric black hole and black brane.

  It is necessary to point out here that we have not studied the
Ramond-Ramond charge fluxes of the black $p$-branes from the method
of anomaly cancelation in this paper. If the scalar field carries
the Ramond-Ramond charge, then it will couple with the Ramond-Ramond
gauge field of the black brane. However, the explicit forms of the
covariant couplings of quantum fields with the Ramond-Ramond gauge
fields are different from that of the $U(1)$ gauge field. At the
same time, the anomaly equations of the Ramond-Ramond charge
currents are also different from that of the electric charge
current. Therefore we cannot use the method of anomaly cancelation
directly for the calculation of the Ramond-Ramond charge fluxes of
the black $p$-branes. For such a problem, we hope to study it in the
future.

\vskip 1cm

\noindent{\Large \bf Acknowledgments}

\vskip 1cm

\noindent The author is grateful for the comments made by the
referee which have improved the contents of this paper.

\vskip 1cm

\noindent{\Large \bf Appendix A. The Conditions for The
Schwarzschild Type Coordinate Transformation}

\vskip 1cm

\noindent  In this appendix, we derive the conditions for the
Schwarzschild type coordinate transformation (30). The original
two-dimensional metric is given by Eq. (28):
$$
  ds^{2}=-A(r)dt^{2}+B(r)dr^{2} ~.
  \eqno{(\mbox{A}.1)}  $$
The unknown coordinate transformation is
$$
  t=t(\tau,r^{\ast}) ~, ~~~~~~~~~~
  r=r(\tau,r^{\ast}) ~.
  \eqno{(\mbox{A}.2)}  $$
Therefore we have
$$
  dt=\frac{\partial t}{\partial \tau}d\tau+
     \frac{\partial t}{\partial r^{\ast}}dr^{\ast} ~,
  \eqno{(\mbox{A}.3)}  $$
$$
  dr=\frac{\partial r}{\partial \tau}d\tau+
     \frac{\partial r}{\partial r^{\ast}}dr^{\ast} ~.
  \eqno{(\mbox{A}.4)}  $$
We demand that under this coordinate transformation, the metric
(A.1) transforms to the form of Eq. (31), i.e.
$$
  ds^{2}=-\sqrt{\frac{A(r^{\ast})}{B(r^{\ast})}}d\tau^{2}+
         \sqrt{\frac{B(r^{\ast})}{A(r^{\ast})}}dr^{\ast 2} ~.
  \eqno{(\mbox{A}.5)}  $$
Here, we mean that the expression for the function $A(r^{\ast})$ is
the same as the expression for the function $A(r)$, the expression
for the function $B(r^{\ast})$ is the same as the expression for the
function $B(r)$.

  To substitute Eqs. (A.3) and (A.4) in Eq. (A.1), then, to compare
with Eq. (A.5), we obtain the following equations:
$$
  -A(r)\left(\frac{\partial t}{\partial \tau}\right)^{2}+
   B(r)\left(\frac{\partial r}{\partial \tau}\right)^{2}=
   -\sqrt{\frac{A(r^{\ast})}{B(r^{\ast})}} ~,
  \eqno{(\mbox{A}.6)}  $$
$$
  -A(r)\left(\frac{\partial t}{\partial r^{\ast}}\right)^{2}+
   B(r)\left(\frac{\partial r}{\partial r^{\ast}}\right)^{2}=
   \sqrt{\frac{B(r^{\ast})}{A(r^{\ast})}} ~,
  \eqno{(\mbox{A}.7)}  $$
$$
  -2A(r)\frac{\partial t}{\partial \tau}
   \frac{\partial t}{\partial r^{\ast}}+
   2B(r)\frac{\partial r}{\partial \tau}
   \frac{\partial r}{\partial r^{\ast}}=0 ~.
  \eqno{(\mbox{A}.8)}  $$
From Eq. (A.4), we can consider that the function $r(\tau,r^{\ast})$
is determined by $\frac{\partial r}{\partial \tau}$ and
$\frac{\partial r}{\partial r^{\ast}}$. Thus, in equations
(A.6)--(A.8), we can regard the function $r(\tau,r^{\ast})$ not as
an independent unknown function. In equations (A.6)--(A.8), the
expressions for the functions $A(r)$, $B(r)$, $A(r^{\ast})$, and
$B(r^{\ast})$ are known, the independent unknown functions are
therefore $\frac{\partial r}{\partial \tau}$, $\frac{\partial
r}{\partial r^{\ast}}$, $\frac{\partial t}{\partial \tau}$, and
$\frac{\partial t}{\partial r^{\ast}}$. Hence there are four
independent unknown functions and three independent equations in
equations (A.6)--(A.8). We can expect that the solutions for
$\frac{\partial r}{\partial \tau}$, $\frac{\partial r}{\partial
r^{\ast}}$, $\frac{\partial t}{\partial \tau}$, and $\frac{\partial
t}{\partial r^{\ast}}$ are existing, and they are not unique
generally. Then, from the solutions of $\frac{\partial r}{\partial
\tau}$, $\frac{\partial r}{\partial r^{\ast}}$, $\frac{\partial
t}{\partial \tau}$, and $\frac{\partial t}{\partial r^{\ast}}$, we
can obtain the solutions of $t(\tau,r^{\ast})$ and
$r(\tau,r^{\ast})$. They are not unique generally. Therefore, such a
kind of coordinate transformation is existing, and it is not unique
generally. However, we need not to obtain its explicit form.

\end{document}